\documentclass[preparint,pra,aps,superscriptaddress,nofootinbib]{revtex4-1}
\usepackage{array}
\usepackage{amsmath}
\usepackage{graphicx}
\usepackage{dsfont}
\usepackage{amssymb}
\usepackage{amsfonts}
\usepackage{indentfirst}
\usepackage{cases}
\usepackage[font=small,labelfont=bf,labelsep=space]{caption}
\makeatletter
\newcommand*{\rom}[1]{\expandafter\@slowromancap\romannumeral
#1@}
\makeatother

\begin{document}
\title{Axiomatic approach for the functional bound of generic Bell's inequality}

\author{Gwangil Bae}
\affiliation{Department of Physics, Sogang University,
Mapo-gu, Shinsu-dong, Seoul 121-742, Korea}

\author{Wonmin Son}
\email{sonwm@physics.org}
\affiliation{Department of Physics, Sogang University,
Mapo-gu, Shinsu-dong, Seoul 121-742, Korea}

\date{\today}

\begin{abstract}
We propose a formalism to derive the maximal bound of generalized Bell type inequalities and shows that the formalism can be applied to various form of Bell functions. The generic Bell function is defined to generate the combinations of all the possible correlations whose local realistic bound can be obtained from the series of the constraint equations. The application of the constraints converts the optimization problem into the counting problems whose complexity is dramatically reduced. It is also shown that generic Bell function can be used to generate many other known Bell type functions such as Mermin, Ardehali, Svetlichny functions for multipartite two-dimensional class.
\end{abstract}
\maketitle

\section{Introduction}
Since J. Bell's monumental discovary to discriminate quantum system from a local realistic model \cite{Bell}, the generalization of the theorem to an arbitrary large quantum system became quite important problem \cite{Clauser69,Mermin80,Gisin92}. The investigation is crucial not just because of the extension of the theorem but also because of the fundamental question whether quantum nonlocality can be an identical notion to quantum entanglement \cite{Peres99}.

Recently, many different approaches in generalization of Bell inequalities (BI) have been introduced. They are graph theoretical approaches \cite{Pitowsky06, Collins04, Sliwa03}, using triangle principle \cite{Kurzynski13}, and quantum correlation \cite{Barrett06, Bancal2011,Bancal12}. Even though the step toward the generalization of Bell inequalities is making a big progress, there are many problems in this category still left unresolved. It is mainly because the complexity in finding the maximal bound of the Bell function is quite high and the number of relevant parameters are getting to be increased exponecially as it is generalized. Historically, the problem had been summaried under the name of ``all the Bell inequality'' and it had been known that it is quite challengeous problem \cite{Peres99}. See also e.g. \cite{Werner01}

The approach using generic correlation have been introduced to identify the general type of Bell function in order to obtaining the full set of Bell's inequalities \cite{Son2006}. In the work, the set of Bell type inequalities for the arbitrary number of high dimensional systems are derived and it is shown that generalised GHZ \cite{Greenberger90,Lee06} violates the generic Bell inequalities maximally. However, the optimization for the local realistic bound was limited to a specified set of correlations and the ultimate formalism for the maximal bound of general correlation is still missing.

In the following, we present aximotized formalism of deriving maximal local realistic bound for the generic Bell functions. The main strategy of our optimization method is to focus on the constraint from the arguments of terms in trigonometrically represented Bell function. Using symmetrical structure in the arguements, one can find the maximal bound of the function by counting the number of independent terms. By applying this formalism to a variant of generic Bell function (GBF), we illustrate the whole process to derive the maximal bound of the function. The generic charater of GBF itself will also be discussed in this paper. We found the form of control factors $\nu$ in GBR with which GBF reduce to ($N$, 2) class Bell type functions derived by D Mermin, G Svetlichny and M Ardehali \cite{Mermin1990, Svetlichny1987, Ardehali1992}.

This paper is organized as following. In the section II, the derivation of maximal bound for the most basic settings of symmetrical Bell inequality, named Clauser-Horn-Shimony-Holt (CHSH) inequality \cite{Clauser69}, has been introduced. The derivation is presented in order to illustrate how the optimization of generalized Bell function have been made. In the section III, we provid the definitions of the general form of the generic Bell function and we present the precise form of the parameters which are appeared in the functions. In the section IV, the maximization procedure of the correlation function has been presented. In the section V, we show that the optimization of the Bell function can be applied to various Bell type correlations and we conclude our paper in the section VI.

\section{Maximal value of Bell correlation with basic settings }
In order to present the basic idea of our approach for the general Bell function in \cite{Son2006}, we start from the most basic settings of the Bell's inequality. The original CHSH version of Bell's inequality is defined in the situation when two parties have two choices of measurements at each site. The two choices of measurements are denoted by $A_i$ and $B_i$ where the subscript $i \in\{1, 2\}$ identifies the measuring parties. In the setting, the Bell function becomes as
\begin{eqnarray}
\langle{\cal B}\rangle= E(a_1, a_2)+E(a_1, b_2)+E(b_1, a_2)-E(b_1, b_2)
=\langle A_1A_2 \rangle+\langle A_1B_2 \rangle+\langle B_1A_2 \rangle-\langle B_1 B_2 \rangle
\end{eqnarray}
where the correlations are given as $E(a_1,a_2)=\langle A_1A_2 \rangle$ and so on. In this case, the local realistic bound can be found rather easily since the average for the measurments $\langle A_1A_2 \rangle=\int d\lambda \rho(\lambda)A_1(\lambda)A_2(\lambda)$ is obtained with the probability distribution $\rho(\lambda)$ of the hidden variable $\lambda$ and measurement values $A_i(\lambda)\in\{1,-1\}$. The bound becomes
\begin{equation}
\langle{\cal B}\rangle=\int d\lambda \rho(\lambda) [A_1(A_2+B_2)+B_1(A_2-B_2)] \le 2
\end{equation}
since $(A_2+B_2)=\pm2$ makes $(A_2-B_2)=0$ and vice versa. Here, we omitted the hidden variable dependence of the measurement parameters $A_i$ and $B_i$. At the same time, the violation of the local realistic bound can be observed when one evaluates the quantum mechanical average of the correlation as
\begin{equation}
E_q(a_1,a_2)=\langle\hat{A}_1\hat{A}_2\rangle=\langle\vec{a}_1\cdot \vec{\sigma}\otimes\vec{a}_2\cdot \vec{\sigma}\rangle=\vec{a}_1\cdot\vec{a}_2=\cos(\theta_1^a-\theta_2^a)
\end{equation}
where $\vec{\sigma}$ is the vector for the Pauli spin operators and the avarage has been taken by one of the maximally entangled state. The famous Cirelson bound is obtained when we consider the full terms
\begin{eqnarray}
\langle\hat{{\cal B}}_q\rangle&=&E_q(a_1,a_2)+E_q(a_1,b_2)+E_q(b_1,a_2)-E_q(b_1,b_2)\\&=&
\cos(\theta_1^a-\theta_2^a)+\cos(\theta_1^a-\theta_2^b)+\cos(\theta_1^b-\theta_2^a)-
\cos(\theta_1^b-\theta_2^b) \le 2\sqrt{2}
\end{eqnarray}
and the upper bound is achieved when $\theta_1^a=0$, $\theta_1^b=\pi/2$, $\theta_2^a=\pi/4$ and  $\theta_2^b=-\pi/4$. With the angles, the values of the correlation functions $E_q(\cdot)$ are $1/\sqrt{2}$ except the last term $E_q(b_1, b_2)$ which is $-1/\sqrt{2}$. The proof that $2\sqrt{2}$ is the maximal value can be made through the evaluation of the norm of Bell operator which had been provided by Cirel'son\cite{Cirelson80}.

Here, we provide the arguement that there can be simpler way to obtain the maximal value for the functional bound of the Bell correlation. The method is quite efficient in general, especially when the number of particles are increased. However, to see how does the optimization works, we consider the simplest two party case which is the case of the CHSH function.The CHSH-Bell correlation can related with the generic form of the Bell function as
\begin{eqnarray}
{\cal B}= CHSH&=& A_1A_2+A_1B_2+B_1A_2-B_1B_2\\
&=&\frac{1-i}{2}(A_1+i B_1)(A_2+i B_2)+c.c.\\
&=& \frac{1}{\sqrt{2}}e^{-i\frac{\pi}{4}}(A_1+i B_1)(A_2+i B_2)+c.c.
\end{eqnarray}
which has the maximal local realistic value $2$. $c.c.$ denotes the complex conjugate of the term in the front. After obtaining the functional form of the correlation, one can symmetrize the correlation through a proper parametrization for the measurement outcomes. The function for the measurement value can be written in a form $A_j=e^{i\pi\alpha_j}$ and $B_j=e^{i\pi\beta_j}$ where $\alpha_j, \beta_j \in\{0,1\}$. With the parameterization, the correlation function will be expanded by symmetertized algebric functions and the symmeterization removes the singularity in the coefficient of the correlation terms as,
\begin{eqnarray}
{\cal B}&=& \frac{1}{\sqrt{2}}e^{-i\frac{\pi}{4}}(e^{i\pi\alpha_1}+i e^{i\pi\beta_1})(e^{i\pi\alpha_2}+i e^{i\pi\beta_2})+c.c.\\
&=&\sqrt{2}\left[\cos\pi(\alpha_1+\alpha_2+1/4)+\cos\pi(\alpha_1+\beta_2+3/4)+\cos\pi(\beta_1+\alpha_2+3/4)
+\cos\pi(\beta_1+\beta_2+5/4)\right].
\end{eqnarray}
Due to the modulo structure, the each term in the correlation function can take the value either $1/\sqrt{2}$ or $-1/\sqrt{2}$ only. Once the value $1/\sqrt{2}$ is assigned in the first three terms (by putting $\alpha_1=\alpha_2=1, \beta_1=\beta_2=0$) the last term becomes $-1/\sqrt{2}$ automatically. It gives maximum local realistic bound $2$ which is true for CHSH inequality. The interesting point in the process of optimization is that the constraint equations with the parameters $\alpha_j$ and $\beta_j$ does not allow all the correlation terms have maximal value $1/\sqrt{2}$. So, in this case, the optimization is achieved by counting the maximal number of terms whose value is assigned by largest number without contradicting the constraint equations.

The quantum mechanical bound for the generic Bell function can be obtained when the measuring values in the correlation function becomes measurement operators. The optimized measurement in this case can be obtained as
\begin{eqnarray}
\hat{{\cal B}}_q&=&\frac{1}{\sqrt{2}}e^{-i\frac{\pi}{4}}(\sigma_x+i \sigma_y)(e^{i\frac{\pi}{4}}\sigma_x+e^{-i\frac{\pi}{4}}\sigma_y)+c.c.
=\frac{1}{\sqrt{2}}(\sigma_x+i \sigma_y)(\sigma_x-i\sigma_y)+c.c.=2\sqrt{2} (\sigma_+\sigma_-+\sigma_-\sigma_+)
\end{eqnarray}
so that the expection value reaches the maximal value $\langle\hat{\cal B}_q\rangle=2\sqrt{2} \langle\phi^+|(\sigma_+\sigma_-+\sigma_-\sigma_+)|\phi^+\rangle=2\sqrt{2}$ where $|\phi^+\rangle=\frac{1}{\sqrt{2}}(|01\rangle+|10\rangle)$. The maximal value of the quantum correlation is given by the largest eignevalue and the value is achieved by one of the eigen states. Thus, the optimization procedure for the local realic bound (and the quantum bound) is algbrically simpler than the usual optimization with the search of the polytop space, especially when the number of particles are increased and the dimensionality of the system becomes higher. It is usually simpler since it is counting problem under the constraint equations rather than numerical optimization.

\section{General correlation for Bell function }
\label{generic}
Generic Bell function is derived by the combinations of all the general correlations in the measurement outcome of multipartite system. Once the combinations of the correlation have been specified, the local realistic bound can be obtained through the algebric optimization procedure. In this section, we provied the prelimiary definition of the general Bell function and derive the functional form of the correlation under the situation of local realistic model.

\subsection{Functional representation of general correlation }
\label{functional}
For our formalism of Bell type cirteria, we need to convert the correlation functions into algebrically analyzable form. By mapping the measurement outcomes in the $d$-root of unity function as $A_j=e^{i2\pi\alpha_j/d}$ and so on, the correlation function can be expressed by the series of the trigonometric function.  Once the correlation function is obtained, inspecting the arguments of trigonometrical terms in the functions, Bell functions can be simply optimized by counting the terms. As an example of our formalism, we will derive maximal bound of generic Bell function (GBF). The trigonometrical representation of the GBF is derived in this section. In the setting of two measurements at the $d$-dimensional $N$-partite systems, the original GBF given in \cite{Son2006} is:
\begin{equation}
\label{Bell_0}
G^{\nu}_N(\lambda)=\frac{1}{2^{N}}\sum_{n=1}^{d-1}
\omega^{\nu n}\prod\limits_{j=1}^N\left(A_j^n+\omega^{\frac{n}{2}}B_j^n\right)+\mbox{c.c.}
\footnote{We used the notation of N-partite GBF as $G^{\nu}_N$ which is oringinally given as $M_{\nu}$
in \cite{Son2006}.}
\end{equation}
\noindent where $\omega = e^{\frac{2\pi i}{d}}$,
$A_j=\omega^{\alpha_j}$, $B_j=\omega^{\beta_j}$ and
c.c. means complex conjugate. The parameters, $\alpha_j$ and
$\beta_j$, representing `the measurement outcomes', are one of the least nonnegative residues modulo $d$;
$0,1,2,\cdot\cdot\cdot, d-1$. 
The indixces `$n$' and
`$j$' denote `high power correlation' and `particle number' respectively. The maximum of the dimension and the parties of a system are denoted by $N$ and $d$ and same convention will be taken throughout this paper. GBF is a Bell type function of arbitrary $N$-partite, $d$-dimensional system. The classical bounds of the function for odd and even $N$ have been found in the previous investigation when $\nu=1/4$ \cite{Son2006}. The mathematical characterization of the GBF is studied later in \cite{Lee2009} more in detail and prove that it is not a tight bound under a linear notation.\footnote{More general correlation function can be considered by taking into account the asymmetric correlations $G^{\nu}_N(\lambda)=\frac{1}{2^{N}}\sum_{n_1=0}^{d-1}\sum_{n_2=0}^{d-1}\cdots
\prod\limits_{j=1}^N\omega^{\nu n_j}\left(A_j^{n_j}+\omega^{{n_j}/2}B_j^{n_j}\right)+\mbox{c.c.}$ but here we simplified the problem by cosidering the correlations of homogeneous power only.} The main purpose of this work is to show explicitly how to derive the maximal bound of the function in detail whose derivation is nonetheless trivial. \footnote{Simple preliminary test given in Appendix \ref{preliminary} shows that Bell theorem cannot be achieved by simply substituting possible maximum value for all cosine terms.}

By mapping $G^{1/4}_N(\lambda)$ into complex plane using $\omega = e^{\frac{2\pi i}{d}}$, $A_j=\omega^{\alpha_j}$, $B_j=\omega^{\beta_j}$, one can have trigonometrical functional form of the GBF consists of cosine terms only,
\begin{equation}
\label{Bell_1}
G^{1/4}_N(\lambda)=\frac{1}{2^{N-1}}\sum_{n=1}^{d-1}\sum_{\gamma=0}^{N}\sum_{k=1}^{\binom{N}{\gamma}}\cos\Big[\frac{n\pi}{2d}\left(
4\mathbb{C}_k^\gamma+2\gamma+1 \right)\Big].
\end{equation}
where the definition of the combinatorics function $\mathbb{C}_k^\gamma$ in the argument will be provided in the following subsection. The combinatorics function is composed of combination of the parameters $\alpha_j$ and $\beta_j$ and the ordering of the parameters is indexed by $\gamma$ and $k$. The functional dependance $G$ by the hidden variable $\lambda$ is determined by the assignment of values to $\alpha_j$ and $\beta_j$. In other word, once the value of $\lambda$ is fixed, the values of $\alpha_j$ and $\beta_j$ are specified.
Another representation of GBF which will be used in the most of our discussion is given by evaluating the summation over dimension $n$ in (\ref{Bell_1}). Arithmetical lemmas for the summation is given in the appendix
A.1. The final form is:
\begin{equation}
\label{Bell_2}
G^{1/4}_N(\lambda)=\frac{1}{2^{N}}\sum_{\gamma=0}^{N}\sum_{k=1}^{\binom{N}{\gamma}}\cot\left[\frac{\pi}{4d}\left(-1\right)^{\gamma}\left(
4\mathbb{C}_k^\gamma+2\gamma+1\right)\right]-1.
\end{equation}
As the effect of summation over $n$, all cosine terms are changed into plus or minus cotangent terms
and an additional `$-1$' terms. The coefficient of each terms are changed as 1/${2^{N}}$ from $1/2^{N-1}$. Equations (\ref{Bell_1}) and (\ref{Bell_2}) gives equivalent trigonometrical representations for a
$\nu=1/4$ which is a variant of GBF defined in \cite{Son2006}.

\subsection{Combination function in the argument}
\label{argument}
The argument of cosine and cotangent in (\ref{Bell_1}) and (\ref{Bell_2}) contains the terms  $4\mathbb{C}_k^\gamma+2\gamma+1$ (neglecting the sign of the argument of cotangent terms).
The term plays important role in the process of functional maximization. Hereafter the argument will be referred to as `the argument function' and simply denoted by $\mathbb{A}^\gamma_k\equiv 4\mathbb{C}_k^\gamma+2\gamma+1$. Necessary conventions related to argument function will be given in the following. The definition of the argument function given here can be also derived by the mapping in Eq. (\ref{Bell_0}).

As we can see from Eq. (\ref{Bell_2}), the variable $\mathbb{C}^\gamma_k$ `the combination function' specifies all the terms in the argument function. Before explaining the explicit form of the combination function, we will define `the combination term' $\mathbb{C}^\gamma$ which does not have $k$ index for the first. The combination term, $\mathbb{C}^\gamma$ is given by the summation of $\gamma$ numbers of $\alpha_i$'s and N-$\gamma$ numbers of $\beta_j$'s without any repetition in their particle number. In other words, $\mathbb{C}^\gamma$ is the summation of N different measurement parameters containing $\gamma$ number of $\beta$ out of all 2N number of measurement parameters;
$\alpha_j$s and $\beta_j$s: $j=1,\,2,\,3,\,\ldots,\, N$. With this definition of $\mathbb{C}^\gamma$, it can be straightforwardly shown that there are $\binom{N}{\gamma}$ number of different choices of combinations for the measurement parameters which correspond to $\mathbb{C}^\gamma$. Those combinations are fully specified by
additional index of $\mathbb{C}^\gamma$, k; `the permutation index'. The index $k$ is determined as one of
$\binom{N}{\gamma}$ positive integers, $1\leq k\leq\binom{N}{\gamma}$. Then we can make one-to-one
correspondence between k and the possible permutation of $\mathbb{C}^\gamma$ for given $\gamma$. We suggests convention for ordering k in Appendix \ref{ordering}. The particular examples of the combination functions can be expanded by measurement parameters as
\begin{align}
\nonumber&\mathbb{C}^0_1 =
\alpha_1+\alpha_2+\alpha_3+\alpha_4+\alpha_5\cdots+\alpha_N
\\\nonumber\\
\nonumber&\mathbb{C}^1_1 =
\beta_1+\alpha_2+\alpha_3+\alpha_4+\alpha_5\cdots+\alpha_N\\
\nonumber&\mathbb{C}^1_2 =
\alpha_1+\beta_2+\alpha_3+\alpha_4+\alpha_5\cdots+\alpha_N\\
\nonumber&\mathbb{C}^1_3 =
\alpha_1+\alpha_2+\beta_3+\alpha_4+\alpha_5\cdots+\alpha_N\\
\nonumber&~~~~~\vdots\quad\quad\quad\quad\quad\quad\quad\quad\vdots\\
\nonumber&\mathbb{C}^1_N =
\alpha_1+\alpha_2+\alpha_3+\alpha_4+\alpha_5\cdots+\beta_N\\
\\
\nonumber&\mathbb{C}^2_1 =
\beta_1+\beta_2+\alpha_3+\alpha_4\cdots+\alpha_{N-1}+\alpha_N\\
\nonumber&\mathbb{C}^2_2 =
\beta_1+\alpha_2+\beta_3+\alpha_4\cdots+\alpha_{N-1}+\alpha_N\\
\nonumber&\mathbb{C}^2_3 =
\beta_1+\alpha_2+\alpha_3+\beta_4\cdots+\alpha_{N-1}+\alpha_N\\
\nonumber&~~~~~\vdots\quad\quad\quad\quad\quad\quad\quad\quad\vdots\\
\nonumber&\mathbb{C}^2_{N(N-1)/2} =
\alpha_1+\alpha_2+\alpha_3+\alpha_4\cdots+\beta_{N-1}+\beta_N.
\end{align}

The argument function is integer because it is a linear summation of integers; $\gamma$ and $\mathbb{C}_k^\gamma$ are nonnegative integers. Also, it is easily verified that the function is one of `odd' integers by its definition $4\mathbb{C}_k^\gamma+2\gamma+1$. Definitions of newly introduced variables in this section; `combination
function', `argument function' are given in Tab.\ref{tab:1}.

\begin{table}[''h'']
\centering
\caption{Definitions of combination function and argument function.}
\begin{tabular}{ >{\centering\arraybackslash} m{3.5cm}
>{\centering\arraybackslash} m{1.5cm}
>{\centering\arraybackslash} m{8cm} }
\hline
Name & Notation & Definition
\\
\hline
combination function & $\mathbb{C}^\gamma_k$ &
\begin{flushleft}
k-th permutation of measurement parameters corresponding to $\mathbb{C}^\gamma$, where the permutation index $k$ is chosen by applying ordering rule Appendix \ref{ordering}, $1\leq k\leq\binom{N}{\gamma}$.
\end{flushleft}
\\
argument function & $\mathbb{A}^\gamma_k$ &
\begin{flushleft}
$4\mathbb{C}_k^\gamma+2\gamma+1$
\end{flushleft}
\\
\hline
\end{tabular}
\label{tab:1}
\end{table}

\subsection{Basic properties of the argument function}
\label{basic}
Before inspecting the algebric structure of generic Bell function (GBF), we identify basic properties of argument function and address about some conventions about the functions defined in the previous section. In order to derive the maximal values of the Bell function (\ref{Bell_2}), it is necessary to count the number of independent terms in argument functions which are appeared in GBF. The properties will be summarized in Theorems \rom{1}, \rom{2} and \rom{3}. Actually the theorems illustrate how to assign the possible values to all the argument functions in section \ref{argument}. So we start with Theorem \rom{1} which is followed by the proof and its corollaries.
\\

\textbf{Theorem \rom{1}.} First two argument functions for $N$-partite system, $\mathbb{A}^0_1$ and $\mathbb{A}^1_k$,
$k=1,2,3, ..., N$ generate all the other argument functions for $N$-partite system. \\

\textbf{Proof.} Let's consider argument functions for $\gamma=0$ and $\gamma=1$;
$\mathbb{A}^0_1=4\mathbb{C}^0_1+1$ and $\mathbb{A}^1_k=4\mathbb{C}^1_k+3$. The possible permutation-ordering  index $k$ for $\mathbb{A}^1_k$ is $1, 2, 3, ..., N$. For given $N$-partite system, those $N+1$ number of argument functions can express all the other argument functions in very brief form. It is because the configuration of measurement parameters in their $\mathbb{C}^0_1$ and $\mathbb{C}^1_k$ are simple:

As we can see, the permutation-ordering index $k$ for $\mathbb{C}^1_k$ is same as the site index of the $\beta$ has. Thus, if we subtract $\mathbb{C}^1_k$ from $\mathbb{C}^0_1$, it gives
\begin{equation}
\label{gen_0}
\mathbb{C}^1_k-\mathbb{C}^0_1=\beta_k-\alpha_k
\end{equation}
Subsequently, it is notable that, by adding (or subtracting) $\mathbb{C}^1_k-\mathbb{C}^0_1=\beta_k-\alpha_k$ to (from) any given $\mathbb{C}^\gamma_k$, any of measurement parameters $\alpha_k$ in $\mathbb{C}^\gamma_k$ can be replaced by $\beta_k$. In this way, arbitrary $\mathbb{C}_k^\gamma$ is given by adding appropriate $\gamma$ number of $\mathbb{C}^1_k-\mathbb{C}^0_1$ to $\mathbb{C}^{\gamma}_k$. For more detailed description, new expression $I^\gamma_k$ is to be introduced. $I^\gamma_k$ is a set of particle site index for all $\beta$'s in $\mathbb{C}^\gamma_k$
\begin{equation}
\label{I}
I^\gamma_k=\left\{ i\mid i \text{ is particle site indices of $\beta$ in $\mathbb{C}^\gamma_k$} \right\}.
\end{equation}
It is clear from its definition that $I^\gamma_k$ as well as $\mathbb{C}^\gamma_k$ is specified by permutation index $k$. In other words, there are one-to-one correspondences between $I^\gamma_k$, $\mathbb{C}^\gamma_k$ and $k$. The relation between $I^\gamma_k$ and $\mathbb{C}^\gamma_k$ for N=4, $\gamma$=2 case is shown in Tab.\ref{tab:1} for example.
\begin{table}[ht]
\centering
\caption{Relation between $\mathbb{C}^2_k$ and $I^2_k$; N=4.}
\begin{tabular}{ >{\centering\arraybackslash} m{1.0cm}>{\centering\arraybackslash}
m{3.5cm}>{\centering\arraybackslash} m{1.0cm} }
\hline
k & $\mathbb{C}^2_k$ & $I^2_k$ \\ [0.5ex]
\hline
1 & $\beta_1+\beta_2+\alpha_3+\alpha_4$ & $\left\{1,2\right\}$
\\
2 & $\beta_1+\alpha_2+\beta_3+\alpha_4$ & $\left\{1,3\right\}$
\\
3 & $\beta_1+\alpha_2+\alpha_3+\beta_4$ & $\left\{1,4\right\}$
\\
4 & $\alpha_1+\beta_2+\beta_3+\alpha_4$ & $\left\{2,3\right\}$
\\
5 & $\alpha_1+\beta_2+\alpha_3+\beta_4$ & $\left\{2,4\right\}$
\\
6 & $\alpha_1+\alpha_2+\beta_3+\beta_4$ & $\left\{3,4\right\}$
\end{tabular}
\label{tab:2}
\end{table}

As a consequence, any $\mathbb{C}^\gamma_k$ can be expressed with $I^\gamma_k$ as
\begin{equation}
\label{gen_3}
\mathbb{C}^\gamma_k=\mathbb{C}^{0}_{1}+\sum_{all\,i}\left(\mathbb{C}^1_i-\mathbb{C}^0_1\right)\quad\left(i\in
I^\gamma_k\right).
\end{equation}
Therefore, $\mathbb{C}^\gamma_k$ can be an arbitrary combination function if the function is for specified measurement outcomes with a measurement choice on the fixed number of parties. The summation over $i$ means the summation over all elements of $I^\gamma_k$. So, this is end of proof that the first two terms of the arguement function can generate any other argument functions. \begin{flushright}
$\Box$
\end{flushright}

As a consequence of the theorem \rom{1}, we can specifies the all the arugment functions $\mathbb{A}^{\gamma}_k$ by fewer numbers of parameters. Let the $N+1$ number of argument functions be congrugent to any constant values $p$ and $q_k$'s upto $4d$ modulus,
\begin{align}
\mathbb{A}^0_1 =4\mathbb{C}^0_1+1&\equiv p \mbox{  (mod 4d)}\nonumber\\
\label{q}
\mathbb{A}^1_k =4\mathbb{C}^1_k+3&\equiv q_k \mbox{  (mod 4d)}
\end{align}
where congrugence relation is expressed by equivalence notation
`$\equiv$'.
Then, with (\ref{q}),
$\mathbb{A}^\gamma_k=4\mathbb{C}_k^\gamma+2\gamma+1$ can be specified as
\begin{eqnarray}
\label{gen_1}
\mathbb{A}^\gamma_k=4\mathbb{C}_k^\gamma+2\gamma+1
&=&4\left\{\mathbb{C}^0_1+\sum_{all\,i}\left(\mathbb{C}^1_i-\mathbb{C}^0_1\right)\right\}+2\gamma+1
=4\left\{\sum_{all\,i}\mathbb{C}^1_i-\left(\gamma-1\right)\mathbb{C}^0_1\right\}+2\gamma+1\\
\nonumber
&\equiv&\sum_{all\,i}\left(\mathbb{A}^1_i-3\right)-\left(\gamma-1\right)\left(\mathbb{A}^0_1-1\right)+2\gamma+1\mbox{
(mod 4d)}\\
\label{gen_2}
&\equiv&\sum_{all\,i}
\mathbb{A}^1_i-\left(\gamma-1\right)\mathbb{A}^0_1\mbox{ (mod
4d)}
\equiv\sum_{all\,i} q_i-\left(\gamma-1\right)p\mbox{ (mod
4d)}\quad\quad\quad\quad\quad\quad\left(i\in I^\gamma_k\right)
\end{eqnarray}
where the relations (\ref{gen_3}) and (\ref{q}) are used in (\ref{gen_1}). Then, with the substitution, we can express arbitrary argument functions as known variables $p$ and $q_k$. It means that the whole set of $q_k$'s ($k=1,2,3, ..., N$) and $p$ can span all the other argument functions and regulates the minimal number of independent parameters for the  constraints of local hidden variable model.\footnote{As a reference, all the equations from (\ref{gen_1}) to (\ref{gen_2}) are folded by $\binom{\gamma}{k}$ number of equations for possible $\binom{\gamma}{k}$ number of different k's. And the moduli of (\ref{gen_1}) and (\ref{gen_2}) are same as that of (\ref{q}) (Theorem A.2.3); same as `$4d$'.
Hereafter, if there is no special reference, all congrugence equations without modulus notations mean $4d$-modulus congrugence equations.} Further statements can be made about the argument functions as following.
\\

\textbf{Corollary \rom{1}-1.} There are $N+1$ independent argument functions for $N$-partite system.\\

\textbf{proof.} The number of first two argument functions is $N+1$ and they are all independent. Thus, Corollary \rom{1}-1. is straightforwardly given from Theorem \rom{1}.
\begin{flushright}
$\Box$
\end{flushright}

\textbf{Corollary \rom{1}-2.} Together with $\mathbb{A}^1_k-\mathbb{A}^0_1$, $\forall k=1, 2, 3, ..., N$, any given argument function for N-partite system generate all the other argument function for N-partite argument function.\\

\textbf{proof.} It will argue that Corollary \rom{1}-2 can be derived from Theorem \rom{1}. First of all, let us denote an arbitrary argument function which is supposed to be fixed in the premise of Corollary \rom{1}-2 as $\mathbb{A}^\gamma_k$. Then $\mathbb{A}^1_k-\mathbb{A}^0_1$, $k=1, 2, 3, ..., N$, and $\mathbb{A}^\gamma_k$ in Corollary\rom{1}-2 can give $\mathbb{A}^0_1$ and $\mathbb{A}^1_k$'s in the supposition of Theorem \rom{1}. We would modify $\mathbb{A}^1_k-\mathbb{A}^0_1$ into terms of measurement parameters as
\begin{eqnarray}
\label{cor_1_2:1}
\mathbb{A}^1_k-\mathbb{A}^0_1 &=&
4\left(\mathbb{C}^1_k-\mathbb{C}^0_1\right)+2
= 4\left(\beta_k-\alpha_k\right)+2.
\end{eqnarray}
The equation is directly given from (\ref{gen_0}). Then using Eq.(\ref{cor_1_2:1}), we have
\begin{eqnarray}
\label{cor_1_2:3}
\mathbb{A}^\gamma_k-\sum_{all\,i}\left(\mathbb{A}^1_i-\mathbb{A}^0_1-2\right)
&=&
\mathbb{A}^\gamma_k-4\sum_{all\,i}\left(\beta_i-\alpha_i\right)
\quad\quad\qquad\left(i\in
I^\gamma_k\right)\nonumber\\
\nonumber &=&
4\left\{\mathbb{C}^\gamma_k-\sum_{all\,i}\left(\beta_i-\alpha_i\right)\right\}+2\gamma+1
\\
\nonumber   &=& 4\mathbb{C}^0_1+2\gamma+1 \\
\label{cor_1_2:4}
&=& \mathbb{A}^0_1
\end{eqnarray}
From the induction, it is found that $\mathbb{A}^0_1$ is determined by $\mathbb{A}^1_k-\mathbb{A}^0_1$'s and $\mathbb{A}^\gamma_k$ which are already known by the supposition of Corollary \rom{1}-2 in (\ref{cor_1_2:4}). Then, adding $\mathbb{A}^0_1$ to $\mathbb{A}^1_k-\mathbb{A}^0_1$, $\mathbb{A}^1_k$'s, $k=1,2,3,..., N$, can be obtained. As $\mathbb{A}^0_1$ and $\mathbb{A}^1_k$'s are given by an arbitrary argument function, Corollary \rom{1}-2 holds by Theorem \rom{1}.
\begin{flushright} $\Box$ \end{flushright}

The properties of argument function which are especially important for our work is given as Theorem
\rom{2} and \rom{3}. Theorem \rom{3} which contains general equation of argument term plays important role for our argument. However, the detailed process of proving them are not so important in our further discussion. So here we presented the results and presented the proofs of them separately in Appendix \ref{proofs} for interested readers.\\

\textbf{Theorem \rom{2}.} All argument functions are determined if all argument functions for two nearest $\gamma$'s;
$4\mathbb{C}_k^t+2t+1$ and $4\mathbb{C}_k^{t-1}+2\left(t-1\right)+1$, $t= 1, 2, 3, ..., N$ are determined.\\

\textbf{Theorem \rom{3}.} If argument terms for two nearest $\gamma$'s are fixed two arbitrary constants Q and
P respectively; $\mathbb{A}^t\equiv Q$ and $\mathbb{A}^{t-1}\equiv P$ for $t=1, 2, 3, ..., N$, then all argument terms for N-partite system are given by arithmetic sequence; $\mathbb{A}^{\gamma\pm z}\equiv \mathbb{A}^\gamma\pm zD$, $\gamma=0, 1, 2, ..., N$ where common difference D is $D\equiv Q-P\equiv q_k-p$, $\forall k=1, 2,
..., \binom{N}{\gamma}$.\\

The general equation for argument function which is used deriving optimal bell function is derived in
Theorem \rom{3}:
\begin{align}\label{general equation}
\mathbb{A}^{\gamma\pm z}\equiv \mathbb{A}^\gamma\pm zD
\end{align}

Every terms of a $\nu=1/4$ variant of GBF is determined by (\ref{general equation}) if argument
functions are reduced to argument term; i.e. terms for same $\gamma$ have same values. It will be proved in the following section that the optimal case satisfies the condition.

\section{Maximal bound for the generic Bell function}
\label{maximal}
The constraints on correlation in generic Bell function are given as properties of argument function in the previous section and they are summerized from Theorem \rom{1} to \rom{3}. They provide sufficient ground for maximizing generic Bell function. Using those constraints, the optimal bounds for the generic Bell function with the control parameter $\nu=1/4$ in Eq. (\ref{Bell_2}) is derived in this section.

First, we derive the constraint equations that make the cotangent terms in Eq. (\ref{Bell_2}) maximum. Given that the argument function can take the odd integer only, it is not difficult to find the possible maximum value of a single cotangent term $\cot{(\pi/4d)}$: see Appendix \ref{preliminary}. Since the argument function is multiplied by $(-1)^\gamma \pi/4d$ in the argument of cotangent term in (\ref{Bell_2}), the individual terms in the correlation are given as $\cot{[(-1)^\gamma(\pi/4d) \mathbb{A}^\gamma_k]}$. In the circumstance, if $\gamma$ is even, maximum value of cotangent term is achieved when the argument function is 1. And for odd $\gamma$, the corresponding maximum condition is given by argument function of $-1$ value. Therefore maximum constraint for cotangent term is reduced as
\begin{equation}
\label{max_con}
\mathbb{A}^\gamma_k=4\mathbb{C}_k^\gamma+2\gamma+1\equiv
\left(-1\right)^{\gamma}\mbox{ (mod 4d)}.
\end{equation}
One now can maximize a cotangent term by assigning constraint (\ref{max_con}) to the argument of the cotangent term. To derive the maximal value of the generic Bell function, we need to apply the maximum constraint  (\ref{max_con}) to the independent argument terms $\mathbb{A}^\gamma_k$ as many as possible. In order to make the GBF in (\ref{Bell_2}) maximized, we need to identify the value of $\gamma$ which is corresponding to the maximum number of combination parameter $k$: $k=\binom{N}{\gamma}$. At the same time,  according to Theorem \rom{3}, the number of independent argument terms is two. Therefore we apply maximum constraint to the two argument terms: $\mathbb{A}^{\gamma_1}$ and $\mathbb{A}^{\gamma_2}$ corresponding to the two most largest combination parameters $\binom{N}{\gamma_1}$ and $\binom{N}{\gamma_2}$ where $\binom{N}{\gamma_1}\geq\binom{N}{\gamma_2}$. If maximum constraint is applied to those two independent argument terms, the values of the others are given by Theorem \rom{3}.\footnote{The detailed mathematical proof of the strategy can be provided and verified through the  upper bounds of  (N,2) class Svetlichny function which will be given by this method in the section V.} Following the procedure of the value assignement, the optimization is reduced to just the counting problem. We will explain the detail procedure in the following for the GBF with $\nu=1/4$.

As shown in equation (\ref{max_con}), the form of maximum constraint depends on the parity of $\gamma$. To obtain the parity of $\gamma_1$ and $\gamma_2$, we use the relation between $N$ and $\gamma$. Because $\gamma_1$ makes maximum number of combination parameter $\binom{N}{\gamma_1}$, $\gamma_1$ is given as $N/2$ and $(N+1)/2$ for even and odd $N$ respectively. However the parity of $\gamma_1$ is not verified by the parity of $N$. For example, let's consider even $N$: $N=2t$ where $t$ is nonnegative integer. Then $\gamma_1$ is given as $N/2=t$ and we do not know the parity of $t$ by our assumption. Similar problem takes place for $\gamma_2$. Therefore we classified the number of party $N$ in 4 modulo values and derive the parity of $\gamma_1$ and $\gamma_2$.

\begin{table}[ht]
\centering
\caption{Determination of the parities of $\gamma_1$ and $\gamma_2$.}
\begin{tabular}{>{\centering\arraybackslash} m{2.0cm}
|>{\centering\arraybackslash} m{3.0cm}
|>{\centering\arraybackslash} m{2.0cm}}
  \hline
  $N$ & $\gamma_1$ & $\gamma_2$ \\ \hline
  $4t$ & even & odd \\ \hline
  $4t+1$ & odd and even & $\cdot$ \\ \hline
  $4t+2$ & odd & even \\ \hline
  $4t+3$ & odd and even & $\cdot$ \\
  \hline
\end{tabular}
\label{tab:3}
\end{table}
$t$ in the first column of Table \rom{3} is one of nonnegative integers. For $N=4t$ case, $\gamma_1=2t$ makes $k$ maximum  and there are two $\gamma_2$ which are both odd as $\gamma_2=2t\pm 1$. For $N=4t+1$, there are two $\gamma_1$ which are $\gamma_1=2t$ and $\gamma_1=2t+1$ which gives same $k=\binom{4t+1}{2t}=\binom{4t+1}{2t+1}$ and we do not need to consider the parity of $\gamma_2$ because the argument terms with maximum $k$ from $\gamma_1$ are doubly folded which are $\mathbb{A}^{2t}$ and $\mathbb{A}^{2t+1}$. Similarly $\gamma_2$ for $N=4t+3$ case is omitted in Table \rom{3}. There are only minor differences in the maximization procedure when the parity of $N$ is same. Moreover the functional bound derived from $G^{1/4}_N$ is same for $N=4t$ and $N=4t+2$ cases. Similary, same bound is derived for $N=4t+1$ and $N=4t+3$. For these reason, without loss of generality, we derived the maximal bounds for odd and even $N$ system using the cases of $N=4t$ and $N=4t+1$ respectively which contain more or less all the cases.

\emph{Maximization of $G^{1/4}_N$ for even $N$ system}: We start with $N=4t$ case as we mentioned in the above. In this case, $\gamma_1=2t$ and $\gamma_2=2t\pm 1$ as we investigated. Hence there are two options for choosing `two' independent argument terms: $\mathbb{A}^{2t}$ and $\mathbb{A}^{2t-1}$ or $\mathbb{A}^{2t}$ and $\mathbb{A}^{2t+1}$. One can show that the maximum of $G^{1/4}_{4t}$ is same for either choices\footnote{One can easily verify it by deriving maximal bounds for both cases.}. Therefore one can derive the bound of $G^{1/4}_{4t}$ by assigning maximum constraint (\ref{max_con}) on $\mathbb{A}^{2t}$ and $\mathbb{A}^{2t-1}$ without loss of generality. The results are
\begin{align}
\nonumber \mathbb{A}^{2t}\equiv  1~~~\mbox{and}~~~
\nonumber \mathbb{A}^{2t-1}\equiv -1.
\end{align}
With this initial condition the common difference $D$ in Theorem \rom{3} is given as $D\equiv \mathbb{A}^{2t}-\mathbb{A}^{2t-1}\equiv 2$. Thus all the terms are given by Theorem \rom{3}.
\begin{align}\label{general equation even}
\mathbb{A}^{\gamma\pm z}&\equiv \mathbb{A}^\gamma\pm 2z
\end{align}
where nonnegative integer $\gamma$ and $z$ satisfy $0\leq \gamma \leq N$ and $0\leq \gamma\pm z\leq N$. Finally, by substituting the argument terms given by equation (\ref{general equation even}) for argument functions in $G^{1/4}_{4t}$: (\ref{Bell_2}), one can derive the maximal bound of $G^{1/4}_{4t}$\footnote{It have to be considered when substituting argument terms for corresponding argument functions that the argument terms are degenerated by $k=\binom{N}{\gamma}$ number of argument functions: see the definition of the argument function in section \ref{basic}.}. The bound of $G^{1/4}_{4t+2}$ is also derived by similar maximization procedure and the result is same as the bound of $G^{1/4}_{4t}$. Consequently the bound of $\nu=1/4$ generic Bell function for even $N$, $G^{1/4}_{N,e}$, is obtained.
\begin{align}
\nonumber
G^{1/4}_{N,e}&\leq\frac{1}{2^N}\left[\sum^{\frac{N}{2}-1}_{z=0}\left(-1\right)^z\left\{\binom{N}{\frac{N}{2}-1-z}+\binom{N}{\frac{N}{2}-z}\right\}\cot\frac{\pi}{4d}\left(2z+1\right)+\left(-1\right)^{\frac{N}{2}}\cot\frac{\pi}{4d}N\right]-1\\
\label{Bell_e}
&=\frac{1}{2^N}\left[\sum^{\frac{N}{2}-1}_{z=0}\left(-1\right)^z\binom{N+1}{\frac{N}{2}-z}\cot\frac{\pi}{4d}\left(2z+1\right)+\left(-1\right)^{\frac{N}{2}}\cot\frac{\pi}{4d}N\right]-1
\end{align}

\emph{Maximization of $G^{1/4}_N$ for odd $N$ system}: For odd $N$ we will consider $N=4t+1$ system. For this case there are two $\gamma_1$: $2t$ and $2t+1$. Accordingly the maximally folded argument terms are $\mathbb{A}^{2t}$ and $\mathbb{A}^{2t+1}$. There is no need to consider the second
most folded argument term given by $\gamma_2$. And the rest of the process is nearly same as that for even $N$ case. The maximum constraints are
\begin{align}
\nonumber \mathbb{A}^{2t}\equiv 1 ~~~\mbox{and}~~~
\nonumber \mathbb{A}^{2t+1}\equiv -1
\end{align}
In this case common difference $D$ is $D\equiv \mathbb{A}^{2t+1}-\mathbb{A}^{2t}\equiv -2$ and the general equation for the argument term is
\begin{align}
\mathbb{A}^{\gamma\pm z}&\equiv \mathbb{A}^\gamma\mp 2z
\end{align}
where integer $\gamma$ and $z$ satisfy $0\leq \gamma \leq N$ and $0\leq \gamma\pm z\leq N$. By similar substitution given in even $N$ case, the maximal bound of $G^{1/4}_{4t+1}$ is derived. And the bound of $G^{1/4}_{4t+3}$ is same as that of $G^{1/4}_{4t+1}$, the bound of $G^{1/4}_N$ for odd $N$ system,  $G^{1/4}_{N,o}$, is given as
\begin{equation}\label{Bell_o}
G^{1/4}_{N,o}\leq\frac{1}{2^{N-1}}\left[\sum^{\frac{N}{2}-1}_{z=0}\left(-1\right)^z\binom{N}{\frac{N-1}{2}-z}\cot\frac{\pi}{4d}\left(2z+1\right)\right]-1
\end{equation}
Therefore, we demonstrated here the whole procedure of maximization for $\nu=1/4$ generic Bell function. The bounds are given in (\ref{Bell_e}) and (\ref{Bell_o}) and they verify the result in \cite{Son2006}. In the following section we generalize $\nu=1/4$ GBF to one that generate multipartite two dimensional Bell type function.

\section{Application to special cases }
We expect that our maximization formalism, derived for $\nu=1/4$, is possible to be extended for other forms of GBF without significant changes in the formalism. As a ground work of the expectation, we derive various form of the multipartite two-dimensional Bell functions, for example, Mermin \cite{Mermin1990, Collins2002}, Svetlichny \cite{Svetlichny1987, Collins2002} and Ardehali \cite{Ardehali1992} functions. The application of our formalism to generic form of Svetlichny function also present in this section.

\subsection{$c/4$ class generic Bell function}
In previous sections we mainly dealt with $\nu=1/4$ generic Bell function. By modifying the forms of $\nu$ in generic Bell function one can derive different variants of GBF. In this section we generalize $\nu$ to any multiples of 1/4 as $\nu=c/4$ where $c$ is an integer. And the periodicity embedded in the structure of $\nu=c/4$ class GBF is investigated. It will be proved in the following subsection that $\nu=c/4$ class GBF generates multipartite two dimensional functions.

Two dimensional generic Bell function with arbitrary $\nu$ is simply derived from (\ref{Bell_0})
\begin{align}\label{gbf_4}
    G^{\nu}_N=\frac{1}{2^N}e^{\nu\pi i}\prod\limits_{j=1}^N(A_j+iB_j)+c.c.
\end{align}
For the coefficient of $A_j$ is unity in the parenthesis, the correlation terms with $\gamma$ number of $B$ measurement have same coefficients containing $i^\gamma=\exp{[\gamma\pi i/2]}$ when neglecting complex conjugate terms. \footnote{In our convention, $\gamma$ is defined as the number of $\beta$ in a correlation term. The definition can be easily modified to the number of $B$ measurement by one-to-one correspondence between $B_j$ and $\beta_j$: $B_j=\omega^{\beta_j}$.} As a result, one can parameterize all $2^N$ number of terms in (\ref{gbf_4}) by $\gamma$: $\gamma=0,\,1,\,2,\,\ldots,\,N$. For the parametrization let's denote the sum of all distinct products of $A$-$B$ measurements including $\gamma$ number of $B$ measurement as `correlation term' $C(\gamma)$, for example $C(0)=A_1A_2\cdots A_N$ and $C(1)=B_1A_2A_3\cdots A_N+A_1B_2A_3A_4\cdots A_N+\cdots+A_1A_2\cdots A_{N-1}B_N$. Expressing $G^{\nu}_N$ with $C(\gamma)$
\begin{align}
\nonumber G^{\nu}_N&=\frac{1}{2^N}e^{\nu\pi i}\sum_{\gamma=0}^{N} \exp{(\gamma\pi i/2)}
C(\gamma)+c.c.\\
\nonumber &=\frac{1}{2^N}e^{\nu\pi i}(1,\,i,\,-1,\,-i,\,\ldots,\,\,\exp{(N\pi i/2)}
)\cdot(C(0),\,C(1),\,C(2),\,\ldots,\,C(N))+c.c.\\
\label{gbf_6}
&=\frac{1}{2^N}e^{\nu\pi i}(1,\,i,\,-1,\,-i,\,\ldots,\,\,\exp{(N\pi i/2)}
)\cdot\vec{C}(N)+c.c.
\end{align}
where `correlation vector' $\vec{C}_N$ is a vector which has $C(\gamma)$ as its $(\gamma+1)$-th element. The expression becomes simpler if $\nu$ is given as a multiple of $1/4$. For the demonstration, the abbreviated expression of $G^{1/4}_N$ is derived by substituting $\nu=1/4$ in (\ref{gbf_6}) and summing the complex conjugate terms.

\begin{align}
\nonumber G^{1/4}_N&=\frac{1}{2^N}e^{\pi i/4}(1,\,i,\,-1,\,-i,\,\ldots,\,\,\exp{(N\pi i/2)})\cdot\vec{C}_N+c.c.\\
\nonumber
&=\frac{1}{2^N\sqrt{2}}(1+i)(1,\,i,\,-1,\,-i,\,\ldots,\,\,\exp{(N\pi i/2)})\cdot\vec{C}_N+c.c.\\\label{gbf_7}
&=\frac{2}{2^{N}\sqrt{2}}(1,\,-1,\,-1,\,1,\,\ldots)\cdot\vec{C}_N\\\label{gbf_8}&
=\frac{1}{2^{N-1/2}}(+,\,-,\,-,\,+,\,\ldots)\cdot\vec{C}_N\end{align}
where `$\ldots$' notation denotes repetition of elements coming before it. In the derivation of (\ref{gbf_7}), coefficients of $C(\gamma)$ are multiplied by 2 by summing the complex
conjugate terms and in equation (\ref{gbf_8}) we abbreviated $\pm1$ as their signs. In general, for $\nu=c/4$ class GBF, one can express the coefficients of $C(\gamma)$ as their signs and a single absolute value. In (\ref{gbf_8}) the absolute values of coefficients of $C(\gamma)$ are same as $1/2^{N-1}\sqrt{2}$. If $\nu\neq c/4$, the absolute values of the real and the imaginary part of phase factor $e^{\nu\pi i}$ in $G^{\nu}_N$ are different. It makes the absolute values of coefficients of $C(\gamma)$ more than one. Therefore one cannot express $G^{\nu}_N$ in simple sign vector form like (\ref{gbf_8}) when $\nu$ is not one of multiples of 1/4.
Finally we define a vector whose $(\gamma+1)$-th element is sign of $C(\gamma)$ in $G^{c/4}_N$ as `sign vector' $\vec{S}^{c/4}_N$. Then $G^{c/4}_N$ can be expressed as an inner product of the sign vector $\vec{S}^{c/4}$ and the correlation vector $\vec{C}_N$.
\begin{subnumcases}{G^{c/4}_N=}\label{vec}
\frac{1}{2^{N-1/2}}\vec{S}^{c/4}_N\cdot\vec{C}_N &\text{($c$ odd)}\label{vec_odd}\\
\frac{1}{2^{N-1}}\vec{S}^{c/4}_N\cdot\vec{C}_N &\text{($c$ even)}\label{vec_even}
\end{subnumcases}
where coefficients of correlation terms are $1/2^{N-1/2}$ and $1/2^{N-1}$ respectively for odd and even $c$. Because of the exponential form of factor $\exp{(c\pi i/4)}$ in $\nu=c/4$ class generic Bell function, there is $8$ modulo structure in $c$: $c=0,\,1,\,2,\,\ldots,\,7$. Equations (\ref{vec_odd}) and (\ref{vec_even}) are verified by deriving vector expression of generic Bell function for according eight forms of $\nu$. The derivations for each cases of $\nu$ are quite similar to that for $\nu=1/4$: see equation (\ref{gbf_8}). Derived sign vectors are given as their period of elements in Table \ref{tab:5}.
\begin{table}[ht]
\centering
\caption{\label{tab:4}The definitions of sign vectors: $\vec{S}^{c/4}_N$; $\nu=c/4$.}
\begin{tabular}{@{}c|cccccccc}
  \hline
  $\nu$ & 0 & 1/4 & 1/2 & 3/4 & 1 & 5/4 & 3/2 & 7/4 \\ \hline
  Sign vector & $(+,0,-,0)$ & $(+,-,-,+)$ & $(0,-,0,+)$ & $(-,-,+,+)$ & $(-,0,+,0)$ & $(-,+,+,-)$ & $(0,+,0,-)$ & $(+,+,-,-)$ \\
  \hline
\end{tabular}
\end{table}

We suggested sign vector notation in this subsection. In Table \ref{tab:4}, we classified all the forms of $\nu=c/4$ class generic Bell function by the period of signs appearing in the structure. With equations (\ref{vec_odd}), (\ref{vec_even}) and Table \ref{tab:4} one can express any $\nu=c/4$ class GBF in the corresponding sign vector form.

\subsection{Derivation of ($N$,2) class Bell type functions}
We derive generic forms of multipartite two dimensional Bell type functions. The periodicities lying in the coefficients of ($N$,2) class Bell functions result in ones that given by sign vectors in Table \ref{tab:4}. One can express ($N$,2) class Bell functions as sign vector form and corresponding $\nu=c/4$ generic Bell function(GBF). The control factors $\nu$ for each of ($N$,2) class Bell functions are found in this section.

As original ($N$,2) class functions we consider Mermin function ($M_N$) \cite{Mermin1990}, Svetlichny function ($S1_3$, $S2_3$) \cite{Svetlichny1987} and Ardehali function ($\mathcal{A}_N$) \cite{Ardehali1992}. Two Svetlichny functions $S1_3$ and $S2_3$ is exceptionally derived for tripartite system. And we also consider Mermin-Collins ($MC_N$) and Svetlichny-Collins ($SC_N$) function recursively derived by Collins \emph{et~al.} in 2002. $SC_N$ is generalized for the multipartite two dimensional system. Without loss of generality, we transformed original definitions which present as an operator forms to the corresponding functional forms which is expressed by dichotomic $A$-$B$ measurements taking $\pm1$ as their outcomes. The functional forms of original functions are given in (\ref{original_functions}).
\begin{subequations}\label{original_functions}
\begin{align}
\label{Mermin}
  M_N =& \frac{1}{2i}\prod_{j=1}^{N}(A_j+iB_j)-\prod_{j=1}^N(A_j-iB_j) \\
\label{Svetlichny_e}
  S1_{3} =& A_1A_2A_3+B_1A_2A_3+A_1B_2A_3+A_1A_2B_3 \\
\nonumber  &-B_1B_2A_3-B_1A_2B_3-A_1B_2B_3-B_1B_2B_3 \\
\label{Svetlichny_o}
  S2_{3} =& A_1A_2A_3-B_1A_2A_3-A_1B_2A_3-A_1A_2B_3 \\
\nonumber  &-B_1B_2A_3-B_1A_2B_3-A_1B_2B_3+B_1B_2B_3 \\
\label{Ardehali}
  \mathcal{A}_N =& \mathcal{A}1+\mathcal{A}2
\end{align}
\end{subequations}
where $\mathcal{A}1$ and $\mathcal{A}2$ is originally defined as operators in Ardehali's paper \cite{Ardehali1992}. Converting operators $\sigma_x^j$ and $\sigma_y^j$ present in \cite{Ardehali1992} to measurements $A_j$ and $B_j$, the functional forms of $\mathcal{A}1$ and $\mathcal{A}2$ are derived. The sign vector forms of $\mathcal{A}1$ and $\mathcal{A}2$ are also easily obtained by inspecting their structures. The sign vector form of the definitions of $\mathcal{A}1$ and $\mathcal{A}2$ are
\begin{subequations}
\begin{align}
 \label{Ardehali1_fist}
    \mathcal{A}1 &=(-,\,0,\,+,\,0,\,-,\,0,\,+,\,\ldots)\cdot\vec{C}_{N-1}\sqrt{2}A_N\\
 \label{Ardehali2_fist}
    \mathcal{A}2 &=(0,\,+,\,0,\,-,\,0,\,+,\,0,\,\ldots)\cdot\vec{C}_{N-1}\sqrt{2}B_N.
\end{align}
\end{subequations}
The functional form of Mermin-Collins, Svetlichny-Collins functions are given in (\ref{Collins_functions}).
\begin{subequations}\label{Collins_functions}
\begin{align}
\label{MC}
  MC_N &= \frac{1}{2}M_{N-1}(A_N+B_N)+\frac{1}{2}M_N'(A_N-B_N) \\
\label{SC_e}
  SC_{N,e} &= M_N, \\
\label{SC_o}
  SC_{N,o} &= \frac{1}{2}(M_N+M'_N)
\end{align}
\end{subequations}

\noindent where subscript $N,e$ and $N,o$ means even and odd N respectively. And $MC_N'$ in (\ref{MC}) and (\ref{SC_o}) is a function given by exchanging $A$ and $B$ measurements in $MC_N$. The initial condition of $MC_N$ is given as $MC_1=A_1$. By inspecting structures of multipartite twodimensional Bell type functions given in (\ref{original_functions}) and (\ref{Collins_functions}), one can find that the sign of correlation terms in those functions are same when the number of $B$ measurement, $\gamma$ is set and that the absolute value of the coefficients are same for all correlation terms. Moreover it is also verified by simple inspection of structure that the period of the signs of correlation terms are found in Table \ref{tab:4}. These constraints in ($N$,2) class Bell functions makes sufficient condition for them to be expressed in their sign vector form. The sign vector expression of functions given in (\ref{original_functions}) and (\ref{Collins_functions}) are respectively given in (\ref{original_sign}) and (\ref{Collins_sign}).
\begin{subequations}\label{original_sign}
\begin{align}
    M_N &= \vec{S}^{-1/2}_N\cdot\vec{C}_N\\
    S1_3 &= \vec{S}^{-1/4}_3\cdot\vec{C}_N\\
    S2_3 &= \vec{S}^{1/4}_3\cdot\vec{C}_N\\
    \mathcal{A}_N &= \sqrt{2}\vec{S}^1\cdot\vec{C}_N
\end{align}
\end{subequations}
\begin{subequations}\label{Collins_sign}
\begin{align}
    MC_N &= 1/2^{N/2}\vec{S}^{(1-N)/4}\cdot\vec{C}_N\\
    SC_{N,e} &= 1/2^{N/2}\vec{S}^{(1-N)/4}\cdot\vec{C}_N\\
    SC_{N,o} &= 1/2^{(N+1)/2}\vec{S}^{-N/4}\cdot\vec{C}_N
\end{align}
\end{subequations}
Now the form of $\nu$ that produce ($N$,2) class Bell type functions are found. Therefore generic form of ($N$,2) class Bell functions are given by (\ref{vec}). The relation between generic Bell function and ($N$,2) class Bell type functions present in Table \ref{tab:5} and Table \ref{tab:6}.
\begin{table}[ht]
\centering
\caption{\label{tab:5} Relation between generic Bell function and original ($N$,2) class functions.}
\begin{tabular}{>{\centering\arraybackslash} m{3.0cm}
|>{\centering\arraybackslash} m{2.0cm}
|>{\centering\arraybackslash} m{2.0cm}|>{\centering\arraybackslash} m{2.0cm}|>{\centering\arraybackslash} m{2.0cm}}
  \hline
  Original functions & $M_N$ & $S1_3$ & $S2_3$ & $A_N$ \\ \hline
  $\nu$ & $-1/2$ & $-1/4$ & 1/4 & 1 \\ \hline
  GBF & $2^{(N-1)}G^{-1/2}_N$ & $2^{5/2}G^{-1/4}_3$ & $2^{5/2}G^{1/4}_3$ & $2^{(N-1/2)}G^1_N$ \\
  \hline
\end{tabular}
\end{table}
\begin{table}[ht]
\centering
\caption{\label{tab:6} Relation between generic Bell function and ($N$,2) class functions generalized by Collins \emph{et~al.}.}
\begin{tabular}{>{\centering\arraybackslash} m{3.5cm}
|>{\centering\arraybackslash} m{3.0cm}
|>{\centering\arraybackslash} m{3.0cm}|>{\centering\arraybackslash} m{3.0cm}}
  \hline
  Generalized functions & $MC_N$ & $SC_{N,e}$ & $SC_{N,o}$ \\ \hline
  $\nu$ & $(1-N)/4$ & $(1-N)/4$ & $-N/4$ \\ \hline
  GBF & $2^{(N-1)/2}G^{(1-N)/4}_N$ & $2^{(N-1)/2}G^{(1-N)/4}_N$ & $2^{(N-2)/2}G^{-N/4}_N$ \\
  \hline
\end{tabular}
\end{table}

Inspecting constraints on correlation terms we proved that $\nu=c/4$ class generic Bell function includes ($N$,2) class Bell type function. Because generic forms of ($N$,2) class Bell type functions are found our optimization formalism given in section \ref{maximal} can be applied with appropriate modification.

\subsection{Maximization of Svetlichny-Collins function}
As an example, we derived functional bound of Svetlichny function in our approach. The constraints on Svetlichny function is quite similar to that on the $\nu=1/4$ generic Bell function given in section \ref{basic}. Accordingly the maximization formalism given in section \ref{maximal} can be applied to Svetlichny-Collins($SC$) function \cite{Collins2002} without significant modification. We generalize the constraints on $\nu=1/4$ GBF to GBF with odd argument function and applied it to Svetlichny function. In this subsection, the functional bounds of Svetlichny function is derived by our optimization formalism.

We preferentially derive maximization constraint of odd argument function which corresponds to that of $\nu=c/4$ GBF: equation (\ref{max_con}). Let's consider cosine representation of generic Bell function for arbitrary $\nu$
\begin{align}
\label{Bell_3}
G^{\nu}_N(\lambda)&=\frac{1}{2^{N-1}}\sum_{n=1}^{d-1}\sum_{\gamma=0}^{N}\sum_{k=1}^{\binom{N}{\gamma}}\cos\Big[\frac{n\pi}{2d}\left(
4\mathbb{C}_k^\gamma+2\gamma+4\nu \right)\Big].
\end{align}
Equation (\ref{Bell_3}) is obtained by mapping $G^\nu_N$: (\ref{Bell_0}) using $\omega = e^{\frac{2\pi i}{d}}$, $A_j=\omega^{\alpha_j}$, $B_j=\omega^{\beta_j}$ like we have done in section \ref{functional} for $\nu=1/4$ GBF. As shown in (\ref{Bell_3}), the form of argument function is changed in GBF with arbitrary $\nu$. Accordingly we redefine the argument function as
\begin{equation}
\label{gen_arg}
\mathbb{A}^\gamma_k=4\mathbb{C}_k^\gamma+2\gamma+4\nu.
\end{equation}
The previous definition of argument function: $4\mathbb{C}_k^\gamma+2\gamma+1$ is $\nu=1/4$ case of (\ref{gen_arg}). The cotangent representation of GBF is derived from (\ref{Bell_3}) by summing in $n$: see Appendix \ref{d-modulus}.
\begin{align}
G^{\nu}_N(\lambda)&=\frac{1}{2^{N}}\sum_{\gamma=0}^{N}\sum_{k=1}^{\binom{N}{\gamma}}\cot\left[\pm\frac{\pi}{4d}\left(
4\mathbb{C}_k^\gamma+2\gamma+4\nu\right)\right]-1.
\label{Bell_4}
\end{align}
We note that in the derivation of (\ref{Bell_4}) we assumed that the value of the argument function is odd. The assumption is valid  for the case of Svetlichny-Collins function whose phase factor $\nu$ in corresponding generic form are $(1-N)/4$ and $-N/4$ respectively for even and odd $N$. It is directly verified by substituting $\nu=(1-N)/4$ and $\nu=-N/4$ to argument function (\ref{gen_arg}) that the argument function is odd for SC function. Because the values of the argument functions are odd by our assumption the values of argument function can be classified into 1 and 3 modulo 4. The plus sign in the argument of cotangent term is for the terms with argument functions satisfying $4\mathbb{C}_k^\gamma+2\gamma+4\nu\equiv 1\;(\mbox{mod 4})$ and the minus sign is for the terms with argument function $4\mathbb{C}_k^\gamma+2\gamma+4\nu\equiv 3\;(\mbox{mod 4})$: see Appendix \ref{d-modulus}. For it is always satisfied that $4\mathbb{C}_k^\gamma\equiv 0\;(\mbox{mod 4})$, the classification depends on the values of $2\gamma$ and $4\nu$; i.e. $4\mathbb{C}_k^\gamma+2\gamma+4\nu\equiv 2\gamma+4\nu\;(\mbox{mod 4})$. $2\gamma$ only can take the value of 0 (mod 4) for even $\gamma$ and 2 (mod 4) otherwise. And $4\nu$ can take values of 1 and 3 (mod 4) because the argument function should be odd under our assumption. All possible summation of $2\gamma$ and $4\nu$ are given in Table \ref{tab:6}.
\begin{table}[ht]
\centering
\caption{The determination of the sign of the cotangent terms.}
\begin{tabular}{>{\centering\arraybackslash} m{2.0cm}
|>{\centering\arraybackslash} m{1.5cm}
|>{\centering\arraybackslash} m{2.5cm}|>{\centering\arraybackslash} m{2.5cm}}
  \hline
  $4\nu$ (mod 4) & $\gamma$ & $2\gamma+4\nu$ (mod 4) & Sign of the cotangent term \\ \hline
  1 & odd & 3 & $-$  \\ \cline{2-4}
   & even & 1 & $+$  \\ \hline
  3 & odd & 1 & $+$  \\ \cline{2-4}
   & even & 3 & $-$  \\ \cline{2-4}
  \hline
\end{tabular}
\label{tab:}
\end{table}

From the results in Table \ref{tab:6}, the functional form of the `$\pm$' sign in (\ref{Bell_4}) is found as $(-1)^{(2\gamma+4\nu-1)/2}$. Therefore exact expression of generic Bell function for odd argument function is given as
\begin{align}
G^{\nu}_N(\lambda)&=\frac{1}{2^{N}}\sum_{\gamma=0}^{N}\sum_{k=1}^{\binom{N}{\gamma}}\cot\left[(-1)^{(\gamma+2\nu-1/2)}\frac{\pi}{4d}\left(
4\mathbb{C}_k^\gamma+2\gamma+4\nu\right)\right]-1.
\label{Bell_5}
\end{align}
Finally one can obtain the maximum condition of the single cotangent term from equation (\ref{Bell_5}). The maximum value of single cotangent term, $\cot{(\pi /4d)}$, is achieved when $\mathbb{A}^\gamma_k\equiv 1$ (mod 4d) for the terms with plus sign, and $\mathbb{A}^\gamma_k\equiv -1$ (mod 4d) otherwise. Therefore the maximum condition for a single cotangent term is given as equation (\ref{max_con_2}).
\begin{align}\label{max_con_2}
A^\gamma_k &\equiv (-1)^{(\gamma+2\nu-1/2)}
\end{align}
Unlike $\nu=1/4$ generic Bell function, one have to consider not only $\gamma$ but also $\nu$ to determine the maximum constraint for a single term in the cotangent form of generic Bell function. If maximum constraint is found, the rest of the maximization procedure is similar for $\nu=1/4$ GBF and generic Bell function with odd argument function. It is because GBF with odd argument function satisfies all constraints given as Theorem \rom{1} to Theorem \rom{3}: see section \ref{basic}. Proofs of the theorems presented in Appendix \ref{proofs} are valid with the generalized definition of the argument function (\ref{gen_arg}). Therefore we briefly explain the optimization procedure for Svetlichny-Collins function. For $SC$ function $\nu$ has form of $\nu=(1-N)/4$ and $\nu=-N/4$ respectively for even and odd $N$. By substituting $\nu$ of $SC$ function into (\ref{max_con_2}), the maximum constraint for $SC$ functions are derived as
\begin{subnumcases}{\mathbb{A}^\gamma_k\equiv}\label{vec}
(-1)^{(\gamma-N/2)} &\text{($N$ even)}\label{max_con_SC}\\
(-1)^{(\gamma-(N-1)/2)} &\text{($N$ odd)}\label{max_con_SC}.
\end{subnumcases}  
Then only left thing to do is to find out $\gamma_1$ and $\gamma_2$, $\gamma$ corresponding to the two most largest combination parameter $k$. After applying maximum condition for the argument function with $\gamma_1$ and $\gamma_2$ using maximum condition (\ref{max_con_SC}) all the other argument functions are determined by Theorem \rom{3}. The functional bounds for Svetlichny's functions: $S_{N,e}$ and $S_{N,o}$ are given in (\ref{S_bound_e}) and (\ref{S_bound_o}).
\begin{align}
\label{S_bound_e}
S_{N,e}&=2^{(N-1)/2}G^{(1-N)/4}_{N,e}\leq\frac{1}{2^{\frac{N-1}{2}}}\left[\sum^{\frac{N}{2}-1}_{z=0}\left(-1\right)^z\binom{N+1}{\frac{N}{2}-z}\cos\frac{\pi}{4}\left(2z+1\right)+\left(-1\right)^{\frac{N}{2}}\cos\frac{\pi}{4}N\right]\\
\label{S_bound_o}
S_{N,o}&=2^{(N-2)/2}G^{-N/4}_{N,o}\leq\frac{1}{2^{\frac{N-2}{2}}}\left[\sum^{\frac{N}{2}-1}_{z=0}\left(-1\right)^z\binom{N}{\frac{N-1}{2}-z}\cos\frac{\pi}{4}\left(2z+1\right)\right]
\end{align}
where we expressed the bounds with cosine functions. It is derived by substituting cotangent terms by cosine terms considering the relation (\ref{cos_cot_relation}) for $n=1$ case. Algebraic calculation of the bounds given in (\ref{S_bound_e}) and (\ref{S_bound_o}) gives 1 for the both. And the results are same as ones that given in \cite{Collins2002}.

In this section, by appropriately generalize the maximum constraint on argument functions, we derived maximum condition for generic Bell function with odd argument function. With the maximum constraint, the bounds of Svetlichny-Collins function is derived from by our optimization formalism.
\section{Conclusion}
In this work, we present aximotized formalism of deriving maximal local realistic bound for the generic Bell functions. The main strategy of our optimization method is to focus on the constraint from the arguments of terms in trigonometrically represented Bell function. Using symmetrical structure in the arguments, one can find the maximal bound of the function by counting the number of independent terms. By applying this formalism to a variant of generic Bell function (GBF), we illustrate the whole process to derive the maximal bound of the function. The generic charater of GBF itself will also be discussed in this paper. We found the form of control factors $\nu$ in GBR with which GBF reduce to ($N$, 2) class Bell type functions. For example, we derived bounds of Svetlichny-Collins(SC) function by applying our optimization formalism to generic form of SC function. The bounds of SC function are derived as 1 and it verifies the result derived by Collins \emph{et~al.} in 2002. It will be interesting work if other possible generalization of our formalism is verified.

\section*{Acknowledgement}
\noindent This work is supported by ICT Research and Development program of MSIP/IITP. (No.2014-044-014-002) and the National Research Foundation of Korea (NRF) grant funded by the Korean Government (No.NRF-2013R1A1A2010537).

\newpage
\appendix

\section{Arithmetics and conventions}
\subsection{Summation over dimension}\label{d-modulus}
Summing dimension in cosine representation of $\nu=1/4$ generic Bell function in (\ref{Bell_1}), the corresponding cotangent form of generic Bell function is derived. Let's preferentially calculate the summation of $\cos{n\theta}$ using arbitrary variable $\theta$.
\begin{equation}
\sum^{d-1}_{n=0}\cos{n\theta}=\mbox{Re}\left[\frac{e^{\imath\theta}\left\{1-e^{\imath\left(d-1\right)\theta}\right\}}{1-e^{\imath\theta}}\right]=\mbox{Re}\left[\frac{e^{\imath\theta}-e^{\imath
d\theta}}{1-e^{\imath\theta}}\right]=\frac{1}{2}\left[\frac{e^{\imath\theta}-e^{\imath
d\theta}}{1-e^{\imath\theta}}-\frac{e^{-\imath\theta}-e^{-\imath
d\theta}}{1-e^{-\imath\theta}}\right]\label{cos_sum_1}
\end{equation}
where we assume that $\theta\neq2\pi m$ where $m$ is integer. For trivial case $\theta=2\pi m$, the summation is simply $d-1$.
Applying Euler's formula: $e^{\imath\theta}=\cos{\theta}+\imath\sin\theta$ to equation (\ref{cos_sum_1}), the summation of $\cos{n\theta}$ is derived as a function of $\theta$: (\ref{cos_sum_2}).
\begin{eqnarray}
\nonumber\sum^{d-1}_{n=0}\cos{n\theta}&=&\frac{1}{2}\left\{\frac{\cos{\left(d-1\right)\theta}-\cos{d\theta}}{1-\cos\theta}-1\right\}=\frac{1}{2}\left\{\frac{\sin{\left(d-\frac{1}{2}\right)\theta}}{\sin{\frac{\theta}{2}}}-1\right\}\\
&=&\frac{1}{2}\left\{\frac{\sin
d\theta\cos\frac{\theta}{2}-\cos
d\theta\sin\frac{\theta}{2}}{\sin\frac{\theta}{2}}-1\right\}=\frac{1}{2}\left\{\sin
d\theta\cot\frac{\theta}{2}-\cos d\theta-1\right\}\label{cos_sum_2}
\end{eqnarray}
Now the general equation (\ref{cos_sum_2}) is classified into four cases given by four different $\theta$: $\theta=\frac{\pi}{2d}\left(4m+n\right)$ where $n=0,\,1,\,2,\,3$.
\begin{eqnarray}
\theta=\frac{\pi}{2d}\left(4m\right)\quad&\rightarrow&\quad\sum^{d-1}_{n=0}\cos{n\theta}=-1\\
\theta=\frac{\pi}{2d}\left(4m+1\right)\quad&\rightarrow&\quad\sum^{d-1}_{n=0}\cos{n\theta}=\frac{1}{2}\left(\cot\frac{\theta}{2}-1\right)\\ \label{cos_mod_1}
\theta=\frac{\pi}{2d}\left(4m+2\right)\quad&\rightarrow&\quad\sum^{d-1}_{n=0}\cos{n\theta}=0\\
\theta=\frac{\pi}{2d}\left(4m+3\right)\quad&\rightarrow&\quad\sum^{d-1}_{n=0}\cos{n\theta}=\frac{1}{2}\left(-\cot\frac{\theta}{2}-1\right) \label{cos_mod_2}
\end{eqnarray}
If $\theta$ is defined as $\theta=\pi(4\mathbb{C}_k^\gamma+2\gamma+1)/2d$, cosine term in equation (\ref{Bell_1}): $\cos{[n\pi(4\mathbb{C}_k^\gamma+2\gamma+1)/2d]}$ are reduced to the form of $\cos{n\theta}$. Because $4\mathbb{C}_k^\gamma+2\gamma+1$ takes value of 1 (mod 4) for even $\gamma$ and 3 (mod 4) for odd, the corresponding cases, (\ref{cos_mod_1}) and (\ref{cos_mod_2}) are the only cases one has to consider. Consequently the summation of cosine function is reduced to cotangent function.
\begin{equation}\label{cos_cot_relation}
\sum^{d-1}_{n=0}\cos{n\theta}=(-1)^\gamma\cot{\frac{\theta}{2}-1}=\cot{\left[(-1)^\gamma \frac{\theta}{2}\right]-1}
\end{equation}


\subsection{Preliminary optimization test of $\nu=1/4$ generic Bell function}
\label{preliminary}
\begin{figure}[h!]
\includegraphics[scale =0.5]{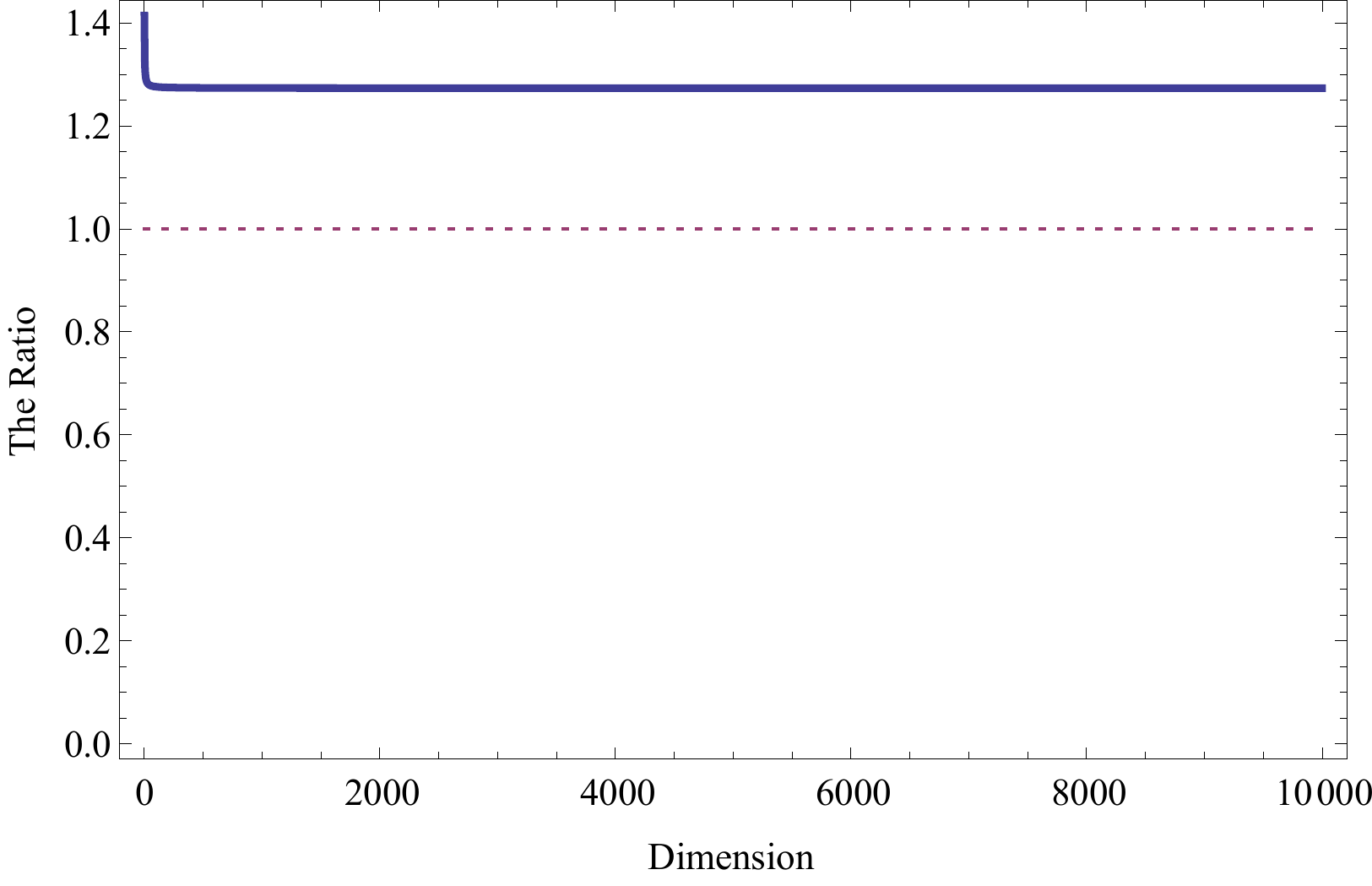}
\caption{Ratio of the trial and quantum bound of $\nu=1/4$ generic Bell function. The ratio converges to $4/\pi$ as $d\rightarrow\infty$.}
\label{diff}
\end{figure}

Here we test the the violation of trigonometrically transformed generic Bell function by neglecting constraints on correlations and assigning maximum value to them. The bound derived by this assumption will be referred to as `trial bound'. It is because trial bound bounds the actual bound of generic Bell function. If the trial bound is violated by quantum mechanical bound, it will not be necessary to derive true maximal bound of generic Bell function(GBF) to verify that GBF satisfies Bell's theorem.   One can find that the argument function $4\mathbb{C}_k^\gamma+2\gamma+1$ in (\ref{Bell_2}) is one of odd integers from its form. Therefore maximum for single cotangent term in (\ref{Bell_2}) is achieved when the argument is `1'. Then if we hypothetically set all $2^N$ number of cotangent terms as their maximum value, $\cot(\pi/4d)$, the trial bound is given as $(1/2^{N})(2^N\cot(\pi/4d))-1=\cot(\pi/4d)-1$.
However as shown in Figure. \ref{diff}, the ratio of the trial bound to quantum bound: $(\cot(\pi/4d)-1)/(d-1)$, is positive for dimension larger than $d=2$.
As a result, one cannot simply conclude that local realistic bound of $\nu=1/4$ generic Bell function is violated by quantum bound in general. Therefore one have to derive true maximal bound of generic Bell function.

\subsection{Ordering combination parameter, k}
\label{ordering}

Ordering rule for combination parameter $k$ of $\mathbb{C}^\gamma_k$ is needed to arrange
$\binom{N}{\gamma}$ number of degenerated combinations of measurement parameters for $\mathbb{C}^\gamma$. Let's consider arbitrary two different combination functions corresponding to $\mathbb{C}^\gamma$. They can be denoted as $\mathbb{C}^\gamma_a$ and $\mathbb{C}^\gamma_b$, $a\neq b$. For $\mathbb{C}^\gamma_a\neq\mathbb{C}^\gamma_b$, there exist particle site index of measurement parameters in $\mathbb{C}^\gamma_a$ and $\mathbb{C}^\gamma_b$ for which the parameters are different in $\mathbb{C}^\gamma_a$ and $\mathbb{C}^\gamma_b$. Let's denote the lowest one of those particle site indices as `low'. Then $\mathbb{C}^\gamma_a$ has $\alpha_{low}$ in its combination and $\mathbb{C}^\gamma_b$ has $\beta_{low}$ or vice versa. Using $\alpha_{low}$ and $\beta_{low}$, we state ordering rule for combination functions for given $\gamma$.\\

\emph{Ordering rule of combinations.} $a>b$, if $\mathbb{C}^\gamma_a$ has
$\alpha_{low}$ and $\mathbb{C}^\gamma_b$ has $\beta_{low}$,
where `low' is the lowest particle site index that gives different
measurement parameter for $\mathbb{C}^\gamma_a$ and
$\mathbb{C}^\gamma_b$.\\

By applying this rule to all possible pairs of $\mathbb{C}^\gamma_a$ and $\mathbb{C}^\gamma_b$, we can give order for all measurement parameter combinations corresponding to $\mathbb{C}_\gamma$. Example case of $N=4$, $\gamma$=2 is given below.
\begin{eqnarray}
\nonumber\mathbb{C}^2_1 &=& \beta_1+\beta_2+\alpha_3+\alpha_4\\
\nonumber\mathbb{C}^2_2 &=& \beta_1+\alpha_2+\beta_3+\alpha_4\\
\nonumber\mathbb{C}^2_3 &=& \beta_1+\alpha_2+\alpha_3+\beta_4\\
\nonumber\mathbb{C}^2_4 &=& \alpha_1+\beta_2+\beta_3+\alpha_4\\
\nonumber\mathbb{C}^2_5 &=& \alpha_1+\beta_2+\alpha_3+\beta_4\\
\nonumber\mathbb{C}^2_6 &=& \alpha_1+\alpha_2+\beta_3+\beta_4
\end{eqnarray}
Here, for $\mathbb{C}^2_1$ and $\mathbb{C}^2_2$, the first particle site index for which $\mathbb{C}^2_1$ and $\mathbb{C}^2_2$ have different measurement parameter is 2. Therefore $\mathbb{C}^2_1$ which has $\beta_{low}=\beta_2$ in it has lower combination parameter than $\mathbb{C}^2_2$.

\section{Proofs of Theorem \rom{2} and \rom{3}}
\label{proofs}
\subsection{Proof of Theorem \rom{2}}
\label{proof_2}
Let the nearest two $\gamma$'s of combination function be $t$ and $t-1$; $(t=
1,\,2,\,3,\,\ldots,\,N)$. And let's suppose that (\ref{gen_t}) and
(\ref{gen_t-1}) which are given by (\ref{gen_2}) holds for
arbitrary constants $Q_k$ and $P_k$.
\begin{align}
\label{gen_t}
\mathbb{A}^t_k\equiv\,\,&4\mathbb{C}_k^t+2t+1\equiv Q_k \\
\label{gen_t-1}
\mathbb{A}^{t-1}_k\equiv\,\,&4\mathbb{C}_k^{t-1}+2\left(t-1\right)+1\equiv P_k 
\end{align}
Using (\ref{gen_t}) and (\ref{gen_t-1}), it will be proved that
argument functions for two nearest $\gamma$ can span all the
other argument functions. Actually $\mathbb{A}^0_1$ and
$\mathbb{A}^1_k$ in Theorem \rom{1} are t=1 case of
$\mathbb{A}_k^t$ and $\mathbb{A}_k^{t-1}$. It might be necessary
for further argument to refer that (\ref{gen_t}) and
(\ref{gen_t-1}) are degenerated by $k=\binom{N}{\gamma}$. They denotes $\binom{N}{t}$ and
$\binom{N}{t-1}$ number of equations respectively. 
Accordingly the number of constants $Q_k$ and $P_k$ are also
$\binom{N}{t}$ and $\binom{N}{t-1}$ respectively.
 Next step is to choose N equations from (\ref{gen_t})
and (\ref{gen_t-1}) respectively. For (\ref{gen_t}), N
number of equations having combination function that include
$\beta_1$, $\beta_2$, ..., $\beta_N$ respectively will be
chosen. Let's denote the chosen combination functions which has
$\beta_n$: $n=1,\,2,\,\ldots,\,N$ as $\mathbb{C}^t_u$.
For (\ref{gen_t-1}), we can select other N number of
equations containing combination function $\mathbb{C}^{t-1}_v$ such that
$\mathbb{C}^{t-1}_v=\mathbb{C}^t_u-\beta_n+\alpha_n$: $n=1,\,2,\,\ldots,\,N$.
Now for every $\mathbb{C}^{t}_u$ containing
$\beta_n$, there exists corresponding $\mathbb{C}^{t-1}_v$ that
have same combination as $\mathbb{C}^{t}_u$ except that
$\mathbb{C}^{t-1}_v$ includes $\alpha_n$ rather than $\beta_n$.
Rewriting (\ref{gen_t}) and (\ref{gen_t-1}) only for
selected those 2N number of equations, (\ref{gen_t_prop_2}) and (\ref{gen_t-1_prop_2}) are given.
\begin{align}
\label{gen_t_prop_2}
\mathbb{A}^t_u\equiv&\,\,4\mathbb{C}_u^t+2t+1
\equiv Q_u\\
\label{gen_t-1_prop_2}
\mathbb{A}^{t-1}_v\equiv&\,\,4\mathbb{C}_v^{t-1}+2\left(t-1\right)+1
\equiv P_v
\end{align}
As we mentioned in the above (\ref{gen_t_prop_2}) and
(\ref{gen_t-1_prop_2}) are related by (\ref{prop_2:1}).
\begin{equation}\label{prop_2:1}
     \mathbb{C}^t_u-\mathbb{C}^{t-1}_v=\beta_n-\alpha_n
\end{equation}
where $n=1,\,2,\,\ldots,\,N$.
Subtracting (\ref{gen_t-1_prop_2}) from
(\ref{gen_t_prop_2}), $\mathbb{A}^1_n-\mathbb{A}^0_1$ is obtained as (\ref{diff_initial}).
\begin{eqnarray}
\nonumber \mathbb{A}^t_u-\mathbb{A}^{t-1}_v &=&
\left(4\mathbb{C}_u^t+2t+1\right)-\left\{4\mathbb{C}_v^{t-1}+2\left(t-1\right)+1\right\}\\
\label{prop_2:2}
   &=& 4\left(\beta_k-\alpha_k\right)+2 \\
\label{diff_initial}   &=& \mathbb{A}^1_n-\mathbb{A}^0_1
\end{eqnarray}
In equation (\ref{prop_2:2}) is used. 
Then from (\ref{gen_t_prop_2}) and
(\ref{gen_t-1_prop_2}), $\mathbb{A}^1_n-\mathbb{A}^0_1$ is
given as a constant $Q_u-P_v$.
\begin{equation}\label{prop_2:3}
  \mathbb{A}^1_n-\mathbb{A}^0_1 \equiv Q_u-P_v
\end{equation}
Now all $\mathbb{A}^1_n-\mathbb{A}^0_1$ is given. Therefore by Corollary \rom{1}-2., Theorem \rom{2}.
holds.

\subsection{Proof of Theorem \rom{3}}
\label{proof_3}
By the assumption of Theorem \rom{3}., let's assign the
values of $\mathbb{A}^t_k$ and $\mathbb{A}^{t-1}_k$ as single
constants $Q$ and $P$.
\begin{align}
\label{gen_t_prop_3}
\mathbb{A}^t_k\equiv\,\,&4\mathbb{C}_k^t+2t+1
\equiv Q\\
\label{gen_t-1_prop_3}
\mathbb{A}^{t-1}_k\equiv\,\,&4\mathbb{C}_k^{t-1}+2\left(t-1\right)+1
\equiv P
\end{align}
By similar procedure given in Appendix \ref{proof_2},
$\mathbb{A}^1_n-\mathbb{A}^0_1$ is determined as
\begin{equation}\label{proof_3:1}
  \mathbb{A}^1_n-\mathbb{A}^0_1 \equiv Q-P\equiv D.
\end{equation}
where $n=1,\,2,\,\ldots,\,N$ and $D$ is a single constant defined as $D \equiv Q-P$. Then argument function is derived by similar logic used for deriving (\ref{gen_2}).
\begin{eqnarray}
\nonumber \mathbb{A}^\gamma_k &=&
4\mathbb{C}_k^\gamma+2\gamma+1\\
\nonumber
&\equiv& \sum_{all\,i}
\mathbb{A}^1_i-\left(\gamma-1\right)\mathbb{A}^0_1\\
&\equiv&
\label{proof_3:3} \mathbb{A}^0_1+\sum_{all\,i}\left(\mathbb{A}^1_i-\mathbb{A}^0_1\right)\\
\label{proof_3:4}
&\equiv& \mathbb{A}^0_1+D\gamma \quad\qquad\qquad\left(i\in
I^\gamma_k\right)
\end{eqnarray}
In equation (\ref{proof_3:4}), $\gamma D$ is derived by substituting of (\ref{proof_3:3})
into (\ref{proof_3:3}). Here $\gamma$
is the number of elements of $I^\gamma_k$. Now Theorem \rom{3} is verified by (\ref{proof_3:4}). 


\end{document}